\documentclass[letterpaper, 10 pt, conference]{ieeeconf} 
\IEEEoverridecommandlockouts                              
\usepackage{graphicx}
\usepackage[letterpaper, left=0.68in, right=0.68in, top=0.75in, bottom=0.75in]{geometry}

\usepackage{subcaption}

\usepackage{etoolbox}

\usepackage{scalerel}

\usepackage[htt]{hyphenat}

\usepackage{siunitx}

\usepackage{pifont}

\usepackage{amsmath,textcomp}

\usepackage{tikz}

\usepackage{comment}
\allowdisplaybreaks

\usepackage{url}

\usepackage{algpseudocode}

\usepackage{paralist}
\usepackage{amsthm}
\usepackage{color}
\usepackage{stmaryrd}
\usepackage{epsfig}
\usepackage{bm}
\usepackage{listings,lstautogobble}
\usepackage{mathtools}
\usepackage{mathrsfs}
\usepackage{amsfonts}
\usepackage{newpxtext,newpxmath}
\usepackage{xspace}
\usepackage{verbatim}
\usepackage{epstopdf}
\usepackage{parcolumns}

\usepackage[utf8]{inputenc}
\usepackage{calrsfs}
\usepackage{rotating}
\usepackage{booktabs}
\usepackage{multirow}

\newcommand{\mycomment}[1]{}

\usepackage[compact]{titlesec}
\titlespacing{\section}{0pt}{0.25ex}{0.25ex}
\titlespacing{\subsection}{0pt}{0.2ex}{0.2ex}
\titlespacing{\subsubsection}{0pt}{0.1ex}{0.1ex}

\usepackage[T1]{fontenc}

\newtheorem*{example}{Motivating Example}
\newtheorem{definition}{Definition}
 \newtheorem{theorem}{Theorem}
  \newtheorem{lemma}{Lemma} 
\usepackage{algorithm}
  \renewcommand{\footnotesize}{\fontsize{8pt}{11pt}\selectfont}

\DeclareMathAlphabet{\mathcal}{OMS}{cmsy}{m}{n}

\usepackage{hyperref}
\hypersetup{
    colorlinks=true,
    linkcolor=black,
    urlcolor=gray,
}
\usepackage{balance}

\newcommand{\x}{\textbf{x}}

\newcommand{\operator}[1]{{\normalfont \texttt{#1}}}

\newcommand*\xor{\oplus}
\newcommand*\xnor{\odot}

\newcommand*\xorsum[2]{\overset{#2}{\underset{#1}{\xor}}}
\newcommand*\andmul[2]{\overset{#2}{\underset{#1}{\Pi}}}

\makeatletter
\newcommand{\vast}{\bBigg@{4}}
\newcommand{\Vast}{\bBigg@{5}}
\makeatother

\makeatletter
\DeclareRobustCommand{\nand}{\mathbin{\mathpalette\n@and@or\land}}
\DeclareRobustCommand{\nor}{\mathbin{\mathpalette\n@and@or\lor}}

\DeclareRobustCommand{\enand}{\overline{\mathbin{\mathpalette\n@and@or\land}}}
\DeclareRobustCommand{\enor}{\overline{\mathbin{\mathpalette\n@and@or\lor}}}

\newcommand{\n@and@or}[2]{
  \vphantom{#2}
  \ooalign{$\m@th#1#2$\cr\hidewidth$\m@th#1\sim$\hidewidth\cr}
}
\makeatother

\begin{document}
\title{\LARGE \bf
Reachability Analysis Using Constrained Polynomial Logical Zonotopes}
\author{
Ahmad Hafez$^{1}$, Frank J. Jiang$^{2}$, Karl H. Johansson$^{2}$, and Amr Alanwar$^{1}$
\vspace{-5cm}
\thanks{ $^{1}$Technical University of Munich; TUM School of Computation, Information and Technology, Department of Computer Engineering. {\tt\small \{a.hafez, alanwar\}@tum.de}.}
\thanks{ $^{2}$KTH Royal Institute of Technology, Stockholm, Sweden. Division of Decision and Control Systems, EECS. They are also affiliated with Digital Futures. {\tt\small\{frankji, kallej\}@kth.se}.}}

\maketitle
\begin{abstract}
In this paper, we propose reachability analysis using constrained polynomial logical zonotopes.
We perform reachability analysis to compute the set of states that could be reached. To do this, we utilize a recently introduced set representation called polynomial logical zonotopes for performing computationally efficient and exact reachability analysis on logical systems. Notably, polynomial logical zonotopes address the "curse of dimensionality" when analyzing the reachability of logical systems since the set representation can represent $2^h$ binary vectors using $h$ generators. After finishing the reachability analysis, the formal verification involves verifying whether the intersection of the calculated reachable set and the unsafe set is empty or not. Polynomial logical zonotopes lack closure under intersections, prompting the formulation of constrained polynomial logical zonotopes, which preserve the computational efficiency and exactness of polynomial logical zonotopes for reachability analysis while enabling exact intersections. Additionally, an extensive empirical study is presented to demonstrate and validate the advantages of constrained polynomial logical zonotopes. 
\end{abstract}
\begin{keywords}
Reachability analysis, logical zonotopes, formal verification.
\end{keywords}

\section{INTRODUCTION}
Reachability analysis is vital for logical systems, ensuring the avoidance of undesired states. However, it encounters a significant challenge in exhaustive state space exploration, often exhibiting exponential complexity, particularly as the number of state variables increases exponentially. There are various approaches to perform reachability analysis on logical systems. When logical systems are modeled as Boolean Networks or Boolean Control Networks (BCN), the reachability of the system relies on the semi-tensor product~\cite{7454743}. Yet, owing to the constraints imposed by point-wise operations and scaling limitations inherent in semi-tensor products, BCN-based approaches become unmanageable for logical systems with high dimensions~\cite{leifeld2019overview}. Recently introduced for reachability analysis in logical systems, mixed zonotopes~\cite{combastel2022functional} and hybrid zonotopes combined with functional decomposition~\cite{siefert2023reachability}, provide an efficient approach to represent boolean functions while maintaining linear memory complexity for reachable sets over time and linear computational complexity relative to the system dimension. Alternatively, some reachability analysis algorithms utilize Binary Decision Diagrams (BDDs), known for their advantages in compact representation and efficient manipulation. However, they face challenges in representing complex systems, often requiring optimal variable ordering, a co-NP-complete problem.

Logical zonotopes, introduced to tackle computational complexity in reachability analysis on logical systems, are sets formed by XORing a binary vector with generators, which are combinations of other binary vectors. They can represent up to $2^h$ points using $h$ generators. Logical zonotopes can facilitate efficient logical operations within the generator space and, thus, reachability analysis~\cite{alanwar2022logical}. Despite their computational advantages, logical zonotopes lack support for exact ANDing in the generator space, leading to the exploration of polynomial logical zonotopes~\cite{alanwar2023polynomial} as a potential solution. The challenge of achieving exact intersections remains for both polynomial logical zonotopes and logical zonotopes. This study introduces constrained polynomial logical zonotopes, demonstrating their effectiveness in achieving exact intersections. This contribution strengthens the theoretical basis of set representations for binary vectors, particularly in logical operations, and promotes more robust computational frameworks across various applications.
This work's contributions are threefold:

\begin{itemize}
  \item We introduce the formulation of constrained polynomial logical zonotopes and detail the application of both Minkowski and Exact XOR, AND, NOT, XNOR, NAND, OR, and NOR logical operations on constrained polynomial logical zonotopes.
  \item We present and prove the exactness of performing set intersection on two constrained polynomial logical zonotopes.
  \item We evaluate and compare the utility of constrained polynomial logical zonotopes with other set representations on the sets intersection and reachability analysis on a high-dimensional boolean function.
\end{itemize}
Readers can reproduce our results by utilizing our openly accessible library\footnotemark.

\footnotetext{\href{https://github.com/aalanwar/Logical-Zonotope}{https://github.com/aalanwar/Logical-Zonotope}}

The subsequent sections of the paper are arranged in the following manner. In section~\ref{sec:prelim}, the preliminary and problem statement are introduced. In section~\ref{sec:CPLZ}, constrained polynomial logical zonotopes are formulated, and the various operations they can accommodate are specified. We evaluate the proposed set representation in section \ref{sec:eval}. Lastly, in section~\ref{sec:con}, we discuss the potential of both representations and address future prospects.

\section{Problem Statement and Preliminaries}\label{sec:prelim}

This section introduces the notation, preliminary definitions, and problem statement.

\subsection{Notation}
Binary set \{0,1\} and natural numbers are expressed by $\mathbb{B}$ and $\mathbb{N}$, respectively. The symbols $\nand$, $\nor$, and $\xnor$ represent the NAND, NOR, and XNOR operations, respectively. Similarly, the symbols $\xor$, $\neg$, $\lor$, and $\land$ express the XOR, NOT, OR, and AND operations, respectively. To simplify notation, we will henceforth express $x \land y$ as $x \, y$ throughout the rest of this work, acknowledging that this is a slight deviation from standard notation. Matrices are represented by uppercase letters, for example, $G \in \mathbb{B}^{n \times k}$, while sets are indicated by uppercase calligraphic letters, as seen in $\mathcal{Z} \subset \mathbb{B}^{n}$. Vectors and scalars are expressed using lowercase letters, such as $b \in \mathbb{B}^{n }$ with elements. The identity matrix of size $n \times n$ is symbolized as $I_n$. The vector $x \in \mathbb{B}^{n}$ is a binary vector of size $n \times 1$.
\subsection{Preliminaries}

Constrained polynomial logical zonotopes are constructed using the Minkowski XOR operation, which we define as follows.

\begin{definition}[Minkowski XOR~\cite{alanwar2022logical}] Given two sets $\mathcal{L}_1$ and $\mathcal{L}_2$ of binary vectors, the Minkowski XOR is defined between every two points in the two sets as
\begin{align}
 \mathcal{L}_1 \xor \mathcal{L}_2 &= \{z_1 \xor z_2| z_1\in \mathcal{L}_1, z_2 \in \mathcal{L}_2 \}. \label{eq:xordef} 
\end{align}
\end{definition}

A constrained polynomial logical zonotope is a generalization of a polynomial logical zonotope, which is defined next. 

\begin{definition}[Polynomial Logical Zonotope~\cite{alanwar2023polynomial}] Given a point $x\in\mathbb{B}^n$ and $h\in\mathbb{N}$ generator vectors in a generator matrix $G=\begin{bmatrix}g_1,\ldots,g_h\end{bmatrix}\in\mathbb{B}^{n\times h}$ , dependent factors identifier $id\in\mathbb{N}^{1\times p}$, exponent matrix $E\in\mathbb{B}^{p\times h}$, a polynomial logical zonotope is defined as:
\begin{equation}
\begin{aligned}
\mathcal{P}=\bigg\{x\in\mathbb{B}^n\mid x=c\xor\xorsum{i=1}{h}\left(\prod_{k=1}^{p}\alpha_k^{E_{(k,i)}}\right)g_i,
\alpha \in \{0,1\}^p\bigg\}.
\end{aligned}
\end{equation}

$\mathcal{P}=\langle c, G, E, id\rangle$ is the shorthand notation for a polynomial logical zonotope.
\end{definition}
A logical zonotope~\cite{alanwar2022logical, alanwar2023polynomial} is a particular case of polynomial logical zonotopes where $E$ is an identity matrix and without $id$ reducing the shorthand definition to $\mathcal{L}=\langle c, G\rangle$.

In Minkowski logical operations, we'll use the \operator{uniqueID} operator to generate a vector of unique integer identifiers. Conversely, for exact logical operations, we'll employ the \operator{mergeID} operator to combine the identical identifiers for the constrained polynomial logical zonotopes, as outlined in~\cite{alanwar2023polynomial}.

\subsection{Problem Statement}

For a system that has a logical function $f: \mathbb{B}^{n_x} \times \mathbb{B}^{n_u} \rightarrow \mathbb{B}^{n_x}$:
\begin{align}
    x(k+1) = f\big(x(k),u(k)\big)
\label{eq:feq}
\end{align}
where $x(k) \in \mathbb{B}^{n_x}$ and $u(k) \in \mathbb{B}^{n_u}$ are the state and the control input, respectively. The logical function $f$ can be composed of different arrangements of logical operators: $\neg,\xor,\land,\xnor,\lor,\nor,$ and $\nand$. We will use binary sets to represent sets of states and inputs for~\eqref{eq:feq}. The reachable set of a system is defined as follows.

\begin{definition}[Exact Reachable Set~\cite{alanwar2022logical}] \label{def:exactreachF}
Given a set of possible inputs $\,\mathcal{U}_k \subset \mathbb{B}^{n_u}$ and a set of initial states $\mathcal{X}_0 \subset \mathbb{B}^{n_x}$, the exact reachable set $\mathcal{R}_{N}$ of \eqref{eq:feq} after $N$ steps is
\begin{align*}
    \mathcal{R}_{N} = \big\{ &x(N) \in \mathbb{B}^{n_x} \; \big| \; \forall k \in \{0,...,N-1\}: \\
        & x(k+1) = f\big(x(k),u(k)\big), 
        \; x(0) \in \mathcal{X}_0, u(k) \in \mathcal{U}_k \big\}.
\end{align*}
\end{definition}

The goal is to calculate the exact forward reachable sets of the system defined in~\eqref{eq:feq}, using constrained polynomial logical zonotopes to generalize logical zonotopes and polynomial logical zonotopes.
\section{Constrained Polynomial Logical Zonotopes}\label{sec:CPLZ}
In this section, the constrained polynomial logical zonotope is introduced along with its set operations.
\begin{example} \label{ex:mot}
Consider a digital circuit with the following boolean functions:
from~\cite {alanwar2023polynomial} with $B_i \in \mathbb{B}^{10}$ and $U_i \in \mathbb{B}^{10}$, $i =1,2,3$, 
\begin{align}
    B_1(k+1) &= U_1(k) \lor ( B_2(k) \xnor B_1(k)),\\
    B_2(k+1) &= B_2(k) \xnor ( B_1(k) \land U_2(k)),\\   
    B_3(k+1) &= B_3(k) \nand ( U_2(k) \xnor U_3(k)).
\end{align}
Our aim throughout the paper is to conduct reachability  starting from multiple input values (sets of values) for $U_i$ with $i=1,2,3$ and determine the possible intersections of $B_i$ with unsafe sets. \end{example}
Next, we introduce the constrained polynomial logical zonotope.
\begin{definition}[Constrained Polynomial Logical Zonotope] Given a point $c\in\mathbb{B}^n$ and $h\in\mathbb{N}$ generator vectors in a generator matrix $G=\begin{bmatrix}g_1,\ldots,g_h\end{bmatrix}\in\mathbb{B}^{n\times h}$, dependent factors identifier $id\in\mathbb{N}^{1\times p}$, exponent matrix $E\in\mathbb{B}^{p\times h}$, 
a constraint generator matrix $A\in \mathbb{B}^{m\times q}$, a constraint vector $b \in \mathbb{B}^{m}$, and a constraint exponent matrix $R \in \mathbb{B}^{p \times q}$, a constrained polynomial logical zonotope is defined as:
\begin{align}
\mathcal{C}=&\bigg\{x\in\mathbb{B}^n\mid x=c \xor \xorsum{i=1}{h} \bigg(\andmul{k=1}{p}\alpha_k^{E_{(k,i)}}\bigg)g_i \nonumber\\
&\,\,\,\, \bigg|\xorsum{i=1}{q}\bigg({\andmul{k=1}{p}{\alpha_{k}}^{{R}_{k,i}}}\bigg)A_{(.,i)}=b, \alpha\in\{0,1\}^p\bigg\}.
\end{align}
For a constrained polynomial logical zonotope, we use the shorthand notation $\mathcal{C}=\langle c, G, E, A, b, R, id\rangle$.
\end{definition}
\subsection{Set Operations}
In the context of two sets comprising binary vectors, a recurrent necessity arises to execute logical operations between these sets' elements. We propose to have the logical operations occur at the generator space of constrained polynomial logical zonotope instead of directly interacting with the set elements. This section derives closed-form expressions of intersection and logical operations in the generator space of constrained polynomial logical zonotopes.
\subsubsection{Sets Intersection}
We start by showing the intersection between logical zonotopes, followed by the intersection between constrained polynomial logical zonotopes.
\paragraph{Logical Zonotopes Intersections}
The intersection between binary sets using logical zonotopes can only be overapproximated by the AND operation between the logical zonotopes, as shown in the following lemma. 
\begin{lemma}
  Given logical zonotopes $\mathcal{L}_1=\langle c_1, G_1\rangle$, and \\$\mathcal{L}_2=\langle c_2, G_2\rangle$ the intersection is overapproximated by $\mathcal{L}_{\cap} = \langle c_{\land},G_{\land} \rangle$ as follows.
\begin{align}
   \mathcal{L}_1 \cap \mathcal{L}_2 &\subseteq \mathcal{L}_{\land}\, 
   \label{eq:andl}
\end{align}
where
$c_{\land} {=} c_{1} c_{2}$ and 
\begin{align}
G_{\land} {=}\big[ &c_{1} g_{2,1},\dots, c_{1} g_{2,h_2}, c_{2} g_{1,1},\dots, c_{2} g_{1,h_1},\nonumber\\ 
& g_{1,1} g_{2,1}, g_{1,1} g_{2,2}, \dots, g_{1,h_1} g_{2,h_2}\big] \, .  
\end{align}
\end{lemma}
\begin{prof}
  Following the lines of~\cite{alanwar2023distributed}, we apply this to logical zonotopes. For a $z \in \mathcal{L}_1 \cap \mathcal{L}_2$
\begin{align}
z= c_1\xorsum{i=1}{h_1} g_{1,i}\beta_{1,i},\label{eq:z1}\\
z= c_2 \xorsum{i=1}{h_2} g_{2,i}\beta_{2,i}.\label{eq:z2}
\end{align}
ANDing \eqref{eq:z1} and \eqref{eq:z2}\\
\begin{align}
\label{eq:zand}
z= c_{1} c_{2} \xorsum{i=1}{h_{2}} c_{1} g_{2,i} {\beta}_{2,i} \xorsum{i=1}{h_{1}} c_{2} g_{1,i} {\beta}_{1,i} \xorsum{i=1,j=1}{h_{1}, h_{2}} g_{1,i} g_{2,j} {\beta}_{1,i} {\beta}_{2,j}.
\end{align}
By representing the $\beta_{1,i}\beta_{2,j}$ by a new $\beta$ we have an over-approximation as logical zonotopes can't represent the $\beta$'s multiplication. This yields that: $z_1z_2 \in \mathcal{L}_{\land}$ and thus $\mathcal{L}_1  \mathcal{L}_2 \subseteq \mathcal{L}_{\land}$.
\end{prof}
As the proved formula in \eqref{eq:zand} is the same as the ANDing overapproximation for logical zonotopes, then the computational complexity of the logical zonotope overapproximated intersection is  $\mathcal{O}(nh_1h_2)$~\cite{alanwar2022logical, alanwar2023polynomial}.
\paragraph{Constrained Polynomial Logical Zonotopes Intersection} 
Exact Intersection remained an unresolved issue with previous logical zonotopes, leading to overapproximations; on the other hand, the intersection between binary sets using constrained polynomial logical zonotopes is exact because of the addition of the constraints, as shown in the following lemma. 
\begin{lemma} \label{lem:intersection}
Given constrained polynomial logical zonotopes $\mathcal{C}_1=\langle c_1, G_1, E_1, A_1, b_1, R_1, id_1\rangle$, and \\$\mathcal{C}_2=\langle c_2, G_2, E_2, A_2, b_2, R_2, id_2\rangle$ the intersection is computed as follows.
\begin{align}\label{eq:intersection}\nonumber
\mathcal{C}_1\cap\mathcal{C}_2 =&\bigg< c_1,G_1,\begin{bmatrix}E_1\\0\end{bmatrix},\begin{bmatrix}A_1&0&0&0\\0&A_2&0&0\\0&0&G_1&G_2\end{bmatrix},
\begin{bmatrix}b_1\\b_2\\c_1 \xor c_2\end{bmatrix},\\&\begin{bmatrix}R_1&\mathbf{0}&E_1&\mathbf{0}\\\mathbf{0}&R_2&\mathbf{0}&E_2\end{bmatrix},\operator{uniqueID}(p_1+p_2)\bigg>.
\end{align}
\end{lemma}
\begin{prof} Following the lines of~\cite{kochdumper2023constrained}, we limit the factors $\alpha_k$ of $\mathcal{C}_1$ to the values of points that belong to $\mathcal{C}_2$ as follows.
\begin{align}\label{eq:constrain}c_1 \xor \xorsum{i=1}{h_1}\left(\andmul{k=1}{p_1}\alpha_k^{E_{1(k,i)}}\right)G_{1(\cdot,i)}&=c_2 \xor \xorsum{i=1}{h_2}\left(\andmul{k=1}{p_2}\alpha_{p_1+k}^{E_{2(k,i)}}\right)G_{2(\cdot,i)}.\end{align}
Using the self-inverse property of XOR, the constraint~\eqref{eq:constrain} can be rewritten as follows.
\begin{align}\label{eq:conint}\xorsum{i=1}{h_1}\left(\andmul{k=1}{p_1}\alpha_k^{E_{1(k,i)}}\right)G_{1(\cdot,i)}\xor\xorsum{i=1}{h_2}\left(\andmul{k=1}{p_2}\alpha_{p_1+k}^{E_{2(k,i)}}\right)G_{2(\cdot,i)}=c_2\xor c_1.\end{align}
This will add the constraint expressed by \eqref{eq:conint} to the other two constraints defined for the intersecting constraint polynomial logical zonotopes, and the intersection can be described as follows.
\begin{align}
\mathcal{C}_1\cap\mathcal{C}_2 =\nonumber
&\bigg\{c_1\oplus\xorsum{i=1}{h_1}\bigg(\andmul{k=1}\nonumber {p_1}\alpha_k^{E_{(k,i)}}\bigg)G_{1(\cdot,i)} \bigg|\xorsum{i=1}{q_1}\bigg(\andmul{k=1}{p_1}\alpha_k^{R_{(k,i)}}\bigg)A_{1(\cdot,i)}=\\ \nonumber
&\,\,\,\,b_1, \xorsum{i=1}{h_1}\bigg(\andmul{k=1}{p_1}\alpha_k^{E_{(k,i)}}\bigg)G_{1(\cdot,i)}\oplus\xorsum{i=1}{h_1}\bigg(\andmul{k=1}{p_2}\alpha_{p_1+k}^{E_{(k,i)}}\bigg)G_{2(\cdot,i)}=\\\nonumber
&\,\,\,\,c_2\oplus c_1, \xorsum{i=1}{q_2}\bigg(\andmul{k=1}{p_2}\alpha_{p_1+i}^{R_{(k,i)}}\bigg)A_{2(\cdot,i)}=b_2,\\
&\,\,\,\,\alpha_k,\alpha_{p_1+k}\in\{0,1\}\bigg\}.
\end{align}
This leads to the same shorthand notation in \eqref{eq:intersection}.
\end{prof}
The computational complexity of this exact intersection operation is $\mathcal{O}(n+p_1+p_2)$~\cite{alanwar2023polynomial}.
\subsubsection{Minkowski Logical Operations}\label{sec:MinLO}
In the next section, we derive the Minkowski logical operations; we start with Minkowski XOR, AND, NOT, XNOR, NAND, OR, and NOR.
\paragraph{Minkowski XOR}
The Minkowski XOR over the generator domain of a constrained polynomial logical zonotope is carried out as follows.
\begin{lemma}\label{lem:MinXOR}Given two constrained polynomial logical zonotopes
$\mathcal{C}_1=\langle c_1, G_1, E_1, A_1, b_1, R_1, id_1\rangle$, and $\mathcal{C}_2=\langle c_2, G_2, E_2, A_2, b_2, R_2, id_2\rangle$ the Minkowski XOR is computed as follows.
\begin{align} \nonumber
\mathcal{C}_1 \oplus \mathcal{C}_2 =& \bigg< c_1 \oplus c_2,
\begin{bmatrix}
G_1 & G_2
\end{bmatrix}, 
{\begin{bmatrix}
E_1 & \mathbf{0} \\
\mathbf{0} & E_2
\end{bmatrix}}, \begin{bmatrix}
A_1 & \mathbf{0} \\
\mathbf{0} & A_2
\end{bmatrix},\nonumber \\
&\,\,\,\,\,\,{\begin{bmatrix}
b_1 \\
b_2
\end{bmatrix}},
\begin{bmatrix}
R_1 & \mathbf{0} \\
\mathbf{0} & R_2
\end{bmatrix}, \text{\operator{uniqueID}} (p_1+p_2)\bigg>.\label{eq:xor}
\end{align}
\end{lemma}
\begin{prof} 
We denote the right-hand side of \eqref{eq:xor} by $\mathcal{C}_\xor$. To prove \eqref{eq:xor}, we need to prove that $\mathcal{C}_\xor \subseteq \mathcal{C}_1 \xor \mathcal{C}_2 $ and $\mathcal{C}_1 \xor \mathcal{C}_2 \subseteq \mathcal{C}_\xor $. For $x_1 \in \mathcal{C}_1$ and $x_2 \in \mathcal{C}_2$, we have 
\begin{align}\exists{\alpha}_1:x_1= \, & \bigg\{c_1\xor\xorsum{i=1}{\mathrm{h}_1}\left(\andmul{k=1}{p_1}{\alpha}_{1,k}^{E_{1,(k,i)}}\right)g_{1,i} \nonumber\\
&\,\,\,\,\bigg| \xorsum{i=1}{q}\bigg({\andmul{k=1}{p}{\alpha_{1,k}}^{{R}_{1,(k,i)}}}\bigg)A_{1,(.,i)}=b_1\bigg\}.\label{eq:x1} \\
\exists{\alpha}_2:x_2= \, & \bigg\{c_2\xor\xorsum{i=1}{\mathrm{h}_2}\left(\andmul{k=1}{p_2}{\alpha}_{2,k}^{E_{2,(k,i)}}\right)g_{2,i} \nonumber\\
&\,\,\,\, \bigg| \xorsum{i=1}{q}\bigg({\andmul{k=1}{p}{\alpha_{2,k}}^{{R}_{2,(k,i)}}}\bigg)A_{2,(.,i)}=b_2\bigg\}.\label{eq:x2}
\end{align}

 Let  ${\alpha}_{\xor,1:p_{\xor}} {=} \begin{bmatrix} {\alpha}_{1,1:p_{1}}\,, {\alpha}_{2,1:p_{2}} \end{bmatrix}$ with $p_{\xor} {=} p_{1} {+} p_{2}$. As XOR is an associative and commutative gate, we get:
\begin{align}
x_1 \xor x_2= &c_\xor \xor\bigg(\xorsum{i=1}{h_\xor}\bigg(\andmul{k=1}{p_\xor}{\alpha}_{\xor,k}^{E_{\xor,(k,i)}}\bigg)g_{\xor,i}\bigg) \nonumber
\end{align}
with the constraint: \begin{align}
\xorsum{i=1}{q_\xor}\bigg(\andmul{k=1}{p_\xor} \alpha_k^{R_\xor(k, i)}\bigg) A_{\xor(\cdot, i)}= b_\xor, \alpha_k, \in\{0,1\} \nonumber
\end{align}
where $c_{\xor} = c_{1} \xor c_{2}$, $G_{\xor} = \begin{bmatrix} G_{1}\,,\, G_{2} \end{bmatrix}$ with $G_{\xor}{=}\Big[ g_{\xor,1},$  ${\dots} ,g_{\xor,q_{\xor}}\Big]$, $E_{\xor}=\begin{bmatrix} E_{1} & 0 \\ 0 & E_{2} \end{bmatrix}$, $A_{\xor}=\begin{bmatrix} A_{1} & 0 \\ 0 & A_{2} \end{bmatrix}$, $b_\xor=\begin{bmatrix}b_1\\b_2\end{bmatrix}$, $R_{\xor}=\begin{bmatrix} R_{1} & 0 \\ 0 & R_{2} \end{bmatrix}$. Thus, $x_1 \xor x_2 \in \mathcal{C}_{\xor}$ and therefore $\mathcal{C}_1 \xor \mathcal{C}_2 \subseteq \mathcal{C}_{\xor}$. Conversely, let $x_{\xor} \in \mathcal{C}_{\xor}$, then 
 \begin{align*}
 \exists {\alpha}_{\xor} &: x_{\xor} = c_{\xor} \xor \xorsum{i=1}{h_\xor} \Big(\andmul{k=1}{p_\xor} {\alpha}_{\xor,k}^{E_{\xor,(k,i)}} \Big) g_{\xor,i}  \, .
 \end{align*}
  Partitioning ${\alpha}_{\xor,1:p_{\xor}}=\begin{bmatrix}{\alpha}_{1,1:p_{1}}\,,\, {\alpha}_{2,1:p_{2}}\end{bmatrix}$, it follows that there exist $x_1 \in \mathcal{C}_1$ and $x_2 \in \mathcal{C}_2$ such that $x_{\xor} = x_1 \xor x_2$. Therefore, $x_{\xor} \in \mathcal{C}_1 \xor \mathcal{C}_2$ and $ \mathcal{C}_{\xor} \subseteq \mathcal{C}_1  \xor \mathcal{C}_2$.\end{prof} 
\mycomment{
\begin{align}\nonumber
\mathcal{C}_1 \xor \mathcal{C}_2\stackrel{\eqref{eq:xordef}}= & \bigg\{c_1 \xor c_2 \xor \xorsum{i=1}{h_1}\bigg(\andmul{k=1}{p_1} \alpha_k^{E_{1(k, i)}}\bigg) G_{1(\cdot, i)} \xor \\\nonumber
& \xorsum{i=1}{h_2}\bigg(\andmul{k=1}{p_2} \alpha_{p_1+k}^{E_{2(k, i)}}\bigg) G_{2(\cdot, i)} \bigg | \xorsum{i=1}{q_1}\bigg(\andmul{k=1}{p_1} \alpha_k^{R_1(k, i)}\bigg) A_{1(\cdot, i)}=\\\nonumber
& b_1,  \xorsum{i=1}{q_2}\bigg(\andmul{k=1}{p_2} \alpha_{p_1+k}^{R_{2(k, i)}}\bigg) A_{2(\cdot, i)}=b_2, \alpha_k, \alpha_{p_1+k} \in\{0,1\}\bigg\}.
\end{align}
Which proves \eqref{eq:xor}.}
The Minkowski XOR has a computational complexity of $\mathcal O(n + p_1 + p_2)$~\cite{alanwar2023polynomial}.
\paragraph{Minkowski AND}
This section shows the proof of the Minkowski AND for constrained polynomial logical zonotopes.
\begin{lemma}
Given two constrained polynomial logical zonotopes $\mathcal{C}_1=\langle c_1, G_1, E_1, A_1, b_1, R_1, id_1\rangle$, and $\mathcal{C}_2=\langle c_2, G_2, E_2, A_2, b_2, R_2, id_2\rangle$ the Minkowski AND is computed as follows.
\begin{align}\nonumber\label{eq:mand}
\mathcal{C}_\land =& \mathcal{C}_1 \land \mathcal{C}_2 \\\nonumber
=& \bigg\langle c_1 c_2, G_\land, E_\land, \begin{bmatrix}A_1&\mathbf{0}\\\mathbf{0}&A_2\end{bmatrix}, \begin{bmatrix}b_1\\b_2\end{bmatrix}, \begin{bmatrix}R_1&\mathbf{0}\\\mathbf{0}&R_2\end{bmatrix},\\
&\,\,\,\,\,\, {\operator{uniqueID}(p_1+p_2+p_1p_2)}\bigg\rangle,
\end{align}
where $G_\land$ and $E_\land$ are as follows.
\begin{align} G_{\wedge}= & {\left[c_1 g_{2,1}, \ldots, c_1 g_{2, h_2}, c_2 g_{1,1}, \ldots, c_2 g_{1, h_1},\right.} \nonumber \\ &\,\, \left.g_{1,1} g_{2,1}, \ldots, g_{1, h_1} g_{2, h_2}\right],\\
E_{\wedge}=& \Bigg[\begin{bmatrix} 0_{p_1\times1}\\E_{2,(.,1)}\end{bmatrix},\ldots,\begin{bmatrix}0_{p_1\times1}\\E_{2,(.,h_2)}\end{bmatrix},\begin{bmatrix}E_{1,(.,1)}\\0_{p_2\times1}\end{bmatrix},\ldots,\begin{bmatrix}E_{1,(.,h_1)}\\0_{p_2\times1}\end{bmatrix},\nonumber \\
&\,\,\,\,\begin{bmatrix}E_{1,(.,1)}\\E_{2,(.,1)}\end{bmatrix},\ldots,\begin{bmatrix}E_{1,(.,h_1)}\\E_{2,(.,h_2)}\end{bmatrix}\Bigg].\end{align}
\end{lemma}
\mycomment{\begin{prof}
The detailed proof is in the extended version of the paper \cite{hafez2024formal}.
\end{prof}}
\begin{prof}
To prove \eqref{eq:mand} we need to prove that $\mathcal{C}_\land \subseteq \mathcal{C}_1 \land \mathcal{C}_2 $ and $\mathcal{C}_1 \land \mathcal{C}_2 \subseteq \mathcal{C}_\land $. For $x_1 \in \mathcal{C}_1$ and $x_2 \in \mathcal{C}_2$, we have 
\begin{align}\exists{\alpha}_1:x_1= \, & \bigg\{c_1\xor\xorsum{i=1}{\mathrm{h}_1}\left(\andmul{k=1}{p_1}{\alpha}_{1,k}^{E_{1,(k,i)}}\right)g_{1,i} \nonumber\\
&\,\,\,\,\bigg| \xorsum{i=1}{q}\bigg({\andmul{k=1}{p}{\alpha_{1,k}}^{{R}_{1,(k,i)}}}\bigg)A_{1,(.,i)}=b_1\bigg\}.\label{eq:x1} \\
\exists{\alpha}_2:x_2= \, & \bigg\{c_2\xor\xorsum{i=1}{\mathrm{h}_2}\left(\andmul{k=1}{p_2}{\alpha}_{2,k}^{E_{2,(k,i)}}\right)g_{2,i} \nonumber\\
&\,\,\,\,\bigg| \xorsum{i=1}{q}\bigg({\andmul{k=1}{p}{\alpha_{2,k}}^{{R}_{2,(k,i)}}}\bigg)A_{2,(.,i)}=b_2\bigg\}.\label{eq:x2}
\end{align}
ANDing \eqref{eq:x1}, \eqref{eq:x2} results in the following.
\begin{align}\nonumber
x_1 x_2= &c_1 c_2\xor\bigg(\xorsum{i=1}{h_2}\bigg(\andmul{k=1}{p_2}{\alpha}_{2,k}^{E_{2,(k,i)}}\bigg)g_{2,i}c_1\bigg) \nonumber\\
&\xor\bigg(\xorsum{i=1}{h_1}\bigg(\andmul{k=1}{p_1}{\alpha}_{1,k}^{E_{1,(k,i)}}g_{1,i}c_2\bigg) \nonumber\\
&\xor\bigg(\xorsum{i_1=1,i_2=1}{{h_1,h_2}}\bigg(\andmul{k_1=1}{p_1}{\alpha}_{1,k_1}^{{E_{1,(k_1,i_1)}}}\bigg)g_{1,i_1}  \bigg(\andmul{k_2=1}{{p_2}}{\alpha}_{2,k_2}^{{E_{2,(k_2,i_2)}}}\bigg)g_{2,i_2}\bigg).
\end{align}
The constraints will be combined as for Minkowski XOR in \eqref{eq:xor}.
Combining the factors in
${\alpha}_{\land} {=} \big[ {\alpha}_{1,1:p_1},\,{\alpha}_{1,1:p_2} \big]$ leads to getting $E_{\land}$ and $G_{\land}$.
Thus, $x_1 x_2 \in \mathcal{C}_{\land}$ and therefore $\mathcal{C}_1 \mathcal{C}_2 \subseteq \mathcal{C}_{\land}$. Conversely, let $x_{\land} \in \mathcal{C}_{\land}$, then 
 \begin{align*}
 \exists {\alpha}_{\land} &: x_{\land} = c_{\land} \xor \xorsum{i=1}{h_{\land}} \Big(\andmul{k=1}{p_{\land}} {\alpha}_{\land,k}^{E_{\land,(k,i)}} \Big) g_{\land,i}  \, .
 \end{align*}
  Partitioning ${\alpha}_{\land} {=} \big[ {\alpha}_{1,1:p_1},\,{\alpha}_{1,1:p_2}\big]$,
  it follows that there exist $x_1 \in \mathcal{C}_1$ and $x_2 \in \mathcal{C}_2$ such that $x_{\land} = x_1 x_2$. Therefore, $x_{\land} \in \mathcal{C}_1 \mathcal{C}_2$ and thus $ \mathcal{C}_{\land} \subseteq \mathcal{C}_1   \mathcal{C}_2$.
\end{prof}
The Minkowski AND computational complexity is $\mathcal O(nh_1h_2 + p_1p_2)$~\cite{alanwar2023polynomial}.
\paragraph{Other Minkowski Operations}
Next, we derive the Minkowski NOT ($\neg$), XNOR ($\xnor$), NAND ($\nand$), OR ($\lor$), NOR ($\nor$) operations.
    In light of the XOR gate's truth table $\neg \mathcal{C}=\mathcal{C}\xor 1=\{\x \xor 1 | x \in \mathcal{C}\}$ which inverts each binary vector in $\mathcal{C}$~\cite{alanwar2022logical}, the Minkowski NOT can be computed as follows. 
\begin{align}\label{eq:MinNOT}\neg \mathcal{C}=\langle c \xor 1_{n \times 1}, G, E, A, b, R, id\rangle\end{align}
The Minkowski NOT computational cost is $\mathcal O(n)$ as for polynomial logical zonotopes~\cite{alanwar2023polynomial}. The Minkowski XNOR can be can be computed as follows.
\begin{align}\mathcal{C}_1 \xnor \mathcal{C}_2 =\neg (\mathcal{C}_1 \xor \mathcal{C}_2) \end{align}
and Minkowski XNOR has a computational complexity of $\mathcal O(n + p_1 + p_2)$~\cite{alanwar2023polynomial}.
Minkowski NAND can be computed as follows.
\begin{align}\label{eq:MinNAND}\mathcal{C}_1 \nand \mathcal{C}_2 =\neg (\mathcal{C}_1 \land \mathcal{C}_2) \end{align}
The Minkowski NAND has a computational complexity of $\mathcal O(nh_1h_2 + p_1p_2)$~\cite{alanwar2023polynomial}.
Utilizing the fact that the NAND gate is a universal gate, we can implement the Minkowski OR ($\lor$) and NOR ($\nor$) using the NAND as follows.
\begin{align*}
&\mathcal{C}_{1} \lor \mathcal{C}_{2} =(\neg{\mathcal C}_{1}){\nand}(\neg{\mathcal C}_{2})\\
&\mathcal{C}_{1} \nor \mathcal{C}_{2} =\neg({\mathcal C}_{1}\vee{\mathcal C}_{2})
\end{align*}
The computational complexity of Minkowski OR is $\mathcal O(nh_1h_2 + p_1p_2)$, while for the Minkowski NOR the computational complexity is $\mathcal O(nh_1h_2 + p_1p_2)$~\cite{alanwar2023polynomial}.
\subsubsection{Exact Logical Operations}
Building on the exact logical operations introduced in polynomial logical zonotopes in~\cite{alanwar2023polynomial}, we use the \operator{mergeID} instead of \operator{uniqueID}, as \operator{uniqueID} is used in the case of Minkowski logical operations~\cite{alanwar2023polynomial}. It is worth mentioning the dependency problem~\cite{combastel2020distributed, combastel2022functional,conf:Sparsepolynomialzonotopes}; in simple words, it is a problem that arises when a binary set shows several times in a calculation, resulting in dealing with the recurring binary set independently. A proposed solution to this issue is using an identifier for each factor~\cite{conf:Sparsepolynomialzonotopes}. We start with executing \operator{mergeID} on the two constrained polynomial logical zonotopes to combine identical identifiers for the constrained polynomial logical zonotopes as an essential step to perform exact logical operations. Next, we show the exact XOR($\bar{\xor}$), and exact AND ($\bar{\land}$).
\paragraph{Exact XOR ($\bar{\xor}$)}
The exact XOR is carried out as follows.
\begin{lemma}\label{lem:xor}
  Given two constrained polynomial logical zonotopes with a common id vector as follows. $\mathcal{C}_1=\langle c_1, G_1, E_1, A_1, b_1, R_1, id\rangle$, and $\mathcal{C}_2=\langle c_2, G_2, E_2, A_2, b_2, R_2, id\rangle$, the exact XOR is computed as follows.
\begin{align}\nonumber
\mathcal{C}_1\xor\mathcal{C}_2=&\Bigg\langle c_1\xor c_2,[G_1,G_2],[E_1,E_2],\begin{bmatrix}A_1&\mathbf{0}\\\mathbf{0}&A_2\end{bmatrix},\begin{bmatrix}b_1\\b_2\end{bmatrix},\\ &\,\,\,\,\begin{bmatrix}R_1&\mathbf{0}\\\mathbf{0}&R_2\end{bmatrix},id\Bigg\rangle\label{eq:exxor}.
\end{align}
\end{lemma}
\begin{prof}
To prove \eqref{eq:exxor} we follow the same steps to prove \eqref{eq:xor}, but taking into consideration that we use here \operator{mergeID} operator and not \operator{uniqueID} operator used in the case of Minkowski logical operations~\cite{alanwar2023polynomial,kochdumper2023constrained}. 
\end{prof}
The exact XOR computational complexity is $\mathcal{O}(n + p_1p_2)$~\cite{alanwar2023polynomial}.
\paragraph{Exact AND ($\bar{\land}$)}

The exact AND is carried out as follows.
\begin{lemma}\label{lem:and}
  Given two constrained polynomial logical zonotopes with a common id vector as follows. $\mathcal{C}_1=\langle c_1, G_1, E_1, A_1, b_1, R_1, id\rangle$, and $\mathcal{C}_2=\langle c_2, G_2, E_2, A_2, b_2, R_2, id\rangle$, the exact AND is computed as $\mathcal{C}_{\bar\land}=\langle c_{\bar\land}, G_{\bar\land}, E_{\bar\land}, A_{\bar\land}, b_{\bar\land}, R_{\bar\land}, id\rangle$ where\\
  \begin{align}
  c_{\bar\land}=&c_1 c_2.\\
  G_{\bar{\land}}=&\Big[c_{1}g_{2,1},\ldots,c_{1}g_{2,h_{2}},c_{2}g_{1,1},\ldots,c_{2}g_{1,h_{1}},\nonumber\\
  &\,\,\,\,g_{1,1}g_{2,1},\ldots,g_{1,h_1}g_{2,h_2}\Big],\\
 E_{\bar{\land}}=&\Big[E_{2,(.,1)},\ldots,E_{2,(.,h_2)},E_{1,(.,1)},\ldots,E_{1,(.,h_1)},\nonumber
 \\&\,\,\,\,max(E_{1,(.,1)},E_{2,(.,1)}),\ldots,max(E_{1,(.,h_1)},E_{2,(.,h_2)})\Big].
 \end{align}
\end{lemma}

\begin{prof}
  To prove the exact AND, we follow the same steps as the proof of Minkowski AND and use the \operator{mergeID} instead of \operator{uniqueID}.
\end{prof}
The exact AND computational complexity is $\mathcal{O}(nh_1h_2 + p_1p_2)$~\cite{alanwar2023polynomial}.
Similarly, the exact NOT proof will be the same as the Minkowski NOT proof, and the id will remain the same. As we have exact AND and exact NOT, we can get the exact NAND, and as stated, the NAND is a universal gate for all exact logic operations.

\subsection{Reachability Analysis}

We use the constrained polynomial logical zonotope to have an exact reachability analysis of~\eqref{eq:feq}, defined in Definition~\ref{def:exactreachF}. This is provided in the following theorem.

\begin{theorem}
\label{thm:reachpoly}
Given a logical function $f: \mathbb{B}^{n_x} \times \mathbb{B}^{n_u} \rightarrow \mathbb{B}^{n_x}$ in \eqref{eq:feq} and starting from initial polynomial logical zonotope $\mathcal{R}_0 \subset \mathbb B^{n_x}$ where $x(0) \in \mathcal{R}_0$, then the exact reachable region computed as:
\begin{align}
    \mathcal{R}_{k+1} =  f\big(\mathcal{R}_{k},\mathcal{U}_k\big)
\end{align}\end{theorem}

\begin{prof}
The logical function comprises XOR and NOT operations, and any logical operations formed using NAND. $\forall x(k) \in \mathcal{R}_{k}$ and $u(k) \in \mathcal{U}_k$, Minkowski XOR and NOT can be computed exactly utilizing Lemma~\ref{lem:MinXOR} and \eqref{eq:MinNOT}, Minkowski NAND using \eqref{eq:MinNAND}. Additionally, we can perform exact XOR using Lemma \ref{lem:xor}, exact AND using Lemma \ref{lem:and}, and exact NAND using exact AND and  NOT operations. The universality of the NAND gate indicates that it can serve as the basis for constructing any other logical gates.
\end{prof}
Constrained polynomial logical zonotopes enable the computation of exact intersections between binary sets and unsafe sets. Afterward, a check is conducted in the point domain to determine whether there is an intersection or if the resulting set remains empty. This involves converting from the generator domain to the point domain by considering all possible combinations of the parameters that satisfy the constraint, which is computationally expensive.

\begin{table*}[tbp]
\caption{Execution Time (milliseconds) for logical zonotopes intersection and number of intersection points.} \label{tab:exint}
\vspace{-2mm}
\centering
\normalsize
\begin{tabular}{ccccccc}
\hline
 & \multicolumn{2}{c}{Log. Zonotope} & \multicolumn{2}{c}{Poly. log. Zonotope} & \multicolumn{2}{c}{Con. poly. log Zonotope} \\ \cmidrule(lr){2-3}  \cmidrule(lr){4-5} \cmidrule(lr){6-7} 
Dimensions & Time   & Size & Time   & Size & Time   & Size \\ \hline
5          & 0.436 & 32   &  0.591 & 20     & 0.350  & 4      \\
7          & 0.437 & 128     & 0.668 & 50     & 0.355 & 2      \\
10         & 0.451 & 512     & 0.727 & 183     & 0.374 & 2      \\
\hline
\end{tabular}
\vspace{-2mm}
\end{table*}

\section{Case Studies}\label{sec:eval}
\vspace{0.1in}
We show various practical use cases to clarify the utilization of operating over the generators' domain of constrained polynomial logical zonotopes. We first illustrate the application of constrained polynomial logical zonotopes for performing intersection and then reachability analysis on a Boolean function with a high-dimensional domain. All of the experiments are executed on a processor Intel(R) Xeon(R) CPU E5-1650 v2 @ 3.50GHz (12 CPUs), with 32.0 GB RAM.

\begin{table*}[tbp]
\caption{Execution Time (seconds) for reachability analysis of a Boolean function (*estimated times).}
\label{tab:exectimebool}
\vspace{-2mm}
\centering
\normalsize
\begin{tabular}{c c c c c c c c c}
\toprule
& \multicolumn{2}{c}{Log. Zonotope} & \multicolumn{2}{c}{Poly. log. Zonotope} & \multicolumn{2}{c}{Con. poly. log. Zonotope} & \multicolumn{2}{c}{BDD}  \\
 \cmidrule(lr){2-3}  \cmidrule(lr){4-5} \cmidrule(lr){6-7} \cmidrule(lr){8-9} 
Steps $N$ & Time & Size & Time & Size & Time & Size & Time & Size \\
\midrule
2 & 0.094 & 768 & 0.103 & 211 & 0.113 & 211 & 0.772 & 211 \\
3 & 0.103 & 896 & 0.115 & 580 &  0.124&580&$3.56 \times 10^5$*    & 580 \\
4 & 0.107 & 896 & 0.168 &  580 &0.237 & 580  &  $4.67 \times 10^6$*     &   -  \\
5 & 0.184 & 896 & 2.054 & 580 &2.090 &580   &   $> 10^7$*     & -    \\
\bottomrule
\end{tabular}
\vspace*{-7mm}
\end{table*}

\subsection{Logical Zonotopes Intersection}

As previously mentioned, a significant contribution of this study is the introduction of exact intersections to logical zonotopes. In this experiment, we generated two random logical zonotopes using two random centers and two random generator matrices. Subsequently, we intersected them using overapproximation for both logical zonotopes and polynomial logical zonotopes, while applying an exact intersection for constrained polynomial logical zonotopes.

The results presented in Table~\ref{tab:exint} highlight that constrained polynomial logical zonotopes achieve the fewest intersection points. Additionally, the overapproximation intersection for polynomial logical zonotopes yields fewer points compared to logical zonotopes, attributed to the presence of exact logical operations in polynomial logical zonotopes. Furthermore, the execution time for logical zonotope intersection is observed to be lower than that of polynomial logical zonotopes, consistent with the computational complexity of AND operations for both types, as discussed in \cite{alanwar2023polynomial}. The exact intersection is only possible with constrained polynomial logical zonotopes, and its execution time is lower than that of the overapproximation intersection for logical zonotopes and polynomial logical zonotopes. The findings indicate that the execution time is influenced by the dimensions of the inputs, with higher dimensions correlating with longer execution times.

\subsection{Reachability Analysis on a High-Dimensional Boolean Function}

Consider the Boolean functions defined in the motivating example in section~\ref{sec:CPLZ}. In the context of reachability analysis, we start by assigning sets comprising two potential values to $B_1(0), B_2(0)$, and $B_3(0)$. Subsequently, we gauge the temporal efficiency of the reachability analysis initiated from this initial condition employing BDDs and logical zonotopes. 
In this example, we opt not to compare with the semi-tensor product-based approach of BCNs due to the impracticality of handling the structure matrix for high-dimensional systems. The structure matrix in such approaches expands exponentially with the number of states and inputs~\cite{alanwar2023polynomial}.

The reachability analysis using BDDs for $N>2$ with the provided variable ordering did not conclude within a reasonable timeframe. Consequently, following~\cite{alanwar2023polynomial}, we opted to utilize the average execution time for a single iteration, multiplying this time to estimate the total duration for the reachability analysis. The findings are presented in Table~\ref{tab:exectimebool}. In high-dimensional systems, logical zonotopes yield a substantial overapproximation. Conversely, constrained polynomial logical zonotopes and polynomial logical zonotopes deliver exact reachability analysis with a minimal execution time.

In polynomial scenarios, Minkowski AND operations are feasible with both polynomial logical zonotopes and constrained polynomial logical zonotopes. This results in a saturation of point count (size) at 580 points after three steps. However, in logical zonotopes, where the AND operation is an overapproximation rather than exact, more points are generated compared to polynomial cases. Constrained polynomial logical zonotopes took execution time slightly longer than that of polynomial logical zonotopes.
\section{Conclusion}\label{sec:con}
In this study, we advocate the utilization of constrained polynomial logical zonotopes to extend the applicability of polynomial logical zonotopes for reachability analysis in logical systems. Constrained polynomial logical zonotopes are constructed by adding a constraint to a polynomial logical zonotope, which allows for the exact computation of the intersection. In different use cases, it was shown that constrained polynomial logical zonotopes were able to perform computationally efficient logical operations and exact set intersections. More use cases for constrained polynomial logical zonotope-based reachability analysis and search algorithms will be investigated in future work.

\bibliographystyle{IEEEtran}
{\small
\bibliography{IEEEabrv,ref}
}

\end{document}